\documentclass[class=preprint]{nature}
\usepackage{graphicx}
\graphicspath{{/Figures/}}
\usepackage[usenames,dvipsnames]{color}
\usepackage[colorlinks=true,citecolor=blue,linkcolor=blue,urlcolor=blue]%
{hyperref}
\usepackage[a4paper]{geometry}
\usepackage{fixltx2e}
\usepackage{upgreek}
\usepackage{amsmath}
\usepackage{amssymb}
\usepackage{amsthm}
\usepackage{mathrsfs}
\usepackage{amsbsy}
\usepackage{amstext}
\usepackage{xcolor}
\usepackage{float}
\usepackage{multirow}
\usepackage{wasysym}
\usepackage{mathtools}
\usepackage{subfigure}
\usepackage{mhchem}
\usepackage{amsfonts}
\usepackage{dcolumn}
\usepackage{placeins}
\usepackage{float}
\usepackage{caption}
\usepackage[right]{lineno}
\usepackage{fancyhdr}
\usepackage{bbold,bm}
\usepackage{pdfpages}
\usepackage[utf8]{inputenc}
\usepackage[T1]{fontenc}%
\usepackage{setspace}
\providecommand{\U}[1]{\protect\rule{.1in}{.1in}}
\DeclareCaptionLabelSeparator{vline}{$\boldsymbol{:}$}
\title{Terahertz stimulated parametric downconversion of a magnon mode in an antiferromagnet}
\author{Zhuquan Zhang$^{1*}$, Yu-Che Chien$^1$, Man Tou Wong$^1$, Frank Y. Gao$^{2}$, Zi-Jie Liu$^1$, Xiaoxuan Ma$^3$,  Shixun Cao$^{3*}$, Edoardo Baldini$^{2*}$, and Keith A. Nelson$^{1*}$ \\
	\normalsize{$^1$Department of Chemistry, Massachusetts Institute of Technology, Cambridge, Massachusetts, USA, 02139 }\\
	\normalsize{$^2$Department of Physics, The University of Texas at Austin, Austin, Texas, USA, 78712 } \\
	\normalsize{$^3$Department of Physics, Materials Genome Institute, Institute for Quantum Science and Technology, Shanghai University, Shanghai, 200444, China}\\
	\normalsize{$^{*}$E-mail: kanelson@mit.edu, edoardo.baldini@austin.utexas.edu, sxcao@shu.edu.cn,\\ zhuquan@mit.edu} \\
}
\begin{document}

\maketitle


\section*{Abstract}
In condensed matter systems, interactions between collective modes offer avenues for nonlinear coherent manipulation of coupled excitations and quantum phases. Antiferromagnets, with their inherently coupled magnon modes, provide a promising platform for nonlinear control of microscopic spin waves and macroscopic magnetization. However, nonlinear magnon-magnon interactions have been only partially elaborated, leaving key gaps in the prospects for potential ultrahigh-bandwidth magnonic signal processing. Here, we use a pair of intense terahertz pulses to sequentially excite two distinct coherent magnon modes in an antiferromagnet and find that the magnon mode with a lower frequency undergoes amplification when the higher-frequency mode is driven. We unveil the nonlinear excitation pathways of this stimulated parametric downconversion process by using polarization-selective two-dimensional terahertz spectroscopy. Our work provides fundamental insights into nonlinear magnonics in antiferromagnets, laying the groundwork for forthcoming spintronic and magnonic devices based on nonlinear magnon-magnon interactions.
\newpage

\section*{Main Text}
 
Parametric excitation and amplification are ubiquitous phenomena in nonlinear systems, occurring when a specific parameter is varied to drive or amplify a coupled degree of freedom. One salient manifestation of this principle is the amplification of light through parametric nonlinear optical processes\cite{boyd2008nonlinear}. If a second-order nonlinear optical medium is pumped by a light wave of frequency $\Omega_0$, spontaneous parametric downconversion\cite{louisell1961quantum,burnham1970observation} can be initiated to generate an entangled photon pair at half of the pump frequency (i.e., $\frac{\Omega_0}{2}$), as shown in Fig. 1A. On the other hand, if a signal wave with frequency $\Omega_1$ is introduced alongside the pump wave at frequency $\Omega_2$, the stimulated version of the downconversion process leads to difference frequency generation (DFG) of an idler wave at frequency $\Omega_2-\Omega_1$ (see Fig. 1B). As a result, a special condition arises when the signal and idler waves are indistinguishable and the pump frequency is precisely double the signal frequency (i.e., $\Omega_2=2\Omega_1$). Under these circumstances, as illustrated in Fig. 1C, the nonlinear process results in the amplification of the signal wave, which is referred to as degenerate parametric amplification\cite{raiford1974degenerate}. Such a nonlinear process is a stimulated counterpart of spontaneous parametric downconversion and has been extensively harnessed to attain desired optical functionalities in fields ranging from optical communication \cite{yuen1978optical,radic2008parametric,marhic2015fiber} and quantum optics \cite{milburn1981production,yamamoto1986preparation,eichler2014quantum} to quantum information processing\cite{ourjoumtsev2006generating,bergeal2010phase}.

In magnetically ordered systems, the generation, manipulation, and detection of spin waves, as well as their quanta (magnons), hold paramount promise for emerging information and signal processing platforms based on spintronics and magnonics\cite{pirro2021advances}. Within this framework, controlling nonlinear interactions among magnon modes—a paradigm known as nonlinear magnonics\cite{zheng2023tutorial}—enables the creation of magnonic states beyond what can be achieved in the linear excitation regime\cite{zhang2024terahertz,zhang2024coupling,leenders2024canted}. This concept bears close analogies to nonlinear optics. One manifestation of this parallel is the parametric excitation of magnons in ferromagnets\cite{bracher2017parallel,lisenkov2019magnetoelastic}. Such a nonlinear process involves interactions between a microwave photon and two magnons from a single spin-wave branch but at varying wavevectors (Fig. 1D). In contrast, the presence of high-frequency magnon modes in complex antiferromagnets\cite{nvemec2018antiferromagnetic,rezende2019introduction,han2023coherent,huang2024extreme} appears to be a highly desirable yet challenging candidate for the nonlinear coherent control of magnon-magnon interactions at terahertz (THz) frequencies\cite{li2023terahertz,li2022perspective}. Only very recently has it been realized that THz excitations of two individual magnon modes in a canted antiferromagnet can lead to coherent photon emission at the difference frequency\cite{zhang2024coupling}, as illustrated in Fig. 1E. Such a nonlinear interaction can amplify the lower-frequency magnon coherence if the parametric resonance condition is satisfied (see Fig. 1F). This mechanism differs from previously established examples involving magnon harmonics\cite{lu2017coherent,huang2024extreme}, conversion\cite{zhang2024terahertz,leenders2024canted}, and coupling to phonons\cite{metzger2024magnon}, which all focus on driving nonlinear responses absent in the linear regime. In contrast, the stimulated version of parametric amplification offers a route to selectively enhance weak magnonic signals, making it highly desirable for both fundamental studies and practical applications.

Here, we devise a protocol to investigate the stimulated parametric downconversion of a coherent magnon mode using polarization-selective two-dimensional (2D) THz spectroscopy. This technique allows us to disentangle the nonlinear responses of various origins and provides sensitivity to the polarization of the desired nonlinear signal. Following this approach, we unveil two distinct nonlinear channels for the amplification of a coherent magnon mode in a canted antiferromagnet and discover a crossover from magnonic DFG to degenerate parametric amplification, a unique pathway to achieve tailored magnonic responses enabled by magnon-magnon interactions.

As a testbed, we select the antiferromagnetic insulator ErFeO$_3$, which crystallizes in an orthorhombic perovskite structure with space group \textit{Pbnm}. Below the  N\'{e}el transition temperature at around 643 K and above the spin reorientation transition temperature at 96 K\cite{yamaguchi2013terahertz}, ErFeO$_3$ orders in the $\Gamma_4$ magnetic phase\cite{li2023terahertz}. In this phase, the spins of neighboring Fe ions are antiferromagnetically aligned along the $a$ axis, but slightly canted towards the $c$ axis, giving rise to a weak net magnetization $\mathbf{M}$ along the $c$ axis. Consequently, two distinct magnon modes exist in the THz frequency range, each following specific selection rules. In the linear response regime, the quasi-antiferromagnetic (qAFM) mode, which corresponds to the oscillation of the magnetization amplitude, is driven by THz radiation with $\mathbf{H}_{THz}\parallel c$, while the quasi-ferromagnetic (qFM) mode, which corresponds to the precession of the net magnetization, is excited by THz radiation with $\mathbf{H}_{THz}\perp c$\cite{yamaguchi2013terahertz,li2018observation}. Although more complex selection rules govern nonlinear responses driven by the THz electric or magnetic components, these responses are significantly weaker than the dominant linear responses and are not observed in our experiment, as will be shown later. Moreover, the magnon frequencies are broadly tunable by changing the temperature in the $\Gamma_4$ phase, making ErFeO$_3$ a promising candidate for observing magnonic parametric amplification.

As an initial step, we elucidate the magnon signal by performing time-domain THz spectroscopy measurements on a (010)-cut ErFeO$_3$ crystal at varying temperatures. As depicted in Fig. 2A, the THz pulse is linearly polarized at 45$^{\circ}$ relative to both the $a$ and $c$ axes. A wire-grid polarizer (WGP) is placed after the sample to detect only the polarization of the emitted signal parallel to that of the incident THz pulse. Under these experimental conditions, both qAFM and qFM modes are excited and detected, manifesting in the time domain as two overlapping oscillatory responses following initial peaks from the incident THz field (see Fig. 2B). Fourier transformations of the time-domain signals shown in Fig. 2C reveal the evolution of the magnon frequencies over a broad temperature range. As the sample temperature is cooled from 340 K to 100 K, the qFM mode undergoes notable softening, especially as it approaches the spin reorientation transition temperature. Conversely, the qAFM mode only exhibits a slight hardening as the temperature decreases, mainly due to moving farther from the N\'{e}el transition at a considerably elevated temperature. To achieve a more comprehensive understanding of the magnon frequency shifts, we model the system with a uniform two-spin Hamiltonian, 
\begin{equation}
		\mathcal{H}=nJ\mathbf{S}_{1} \mathbf{S}_{2}+n\mathbf{D}\cdot (\mathbf{S}_{1}\times \mathbf{S}_{2})-\sum_{i=1,2}(K_{a}S_{ia}^2+K_{c}S_{ic}^2),
\end{equation}  
where the first term denotes the antiferromagnetic exchange interaction, the second term represents the Dzyaloshinskii–Moriya interaction leading to spin canting, and the third term describes the magnetocrystalline anisotropy energy. Here, $\mathbf{S}_{1}$ and $\mathbf{S}_{2}$ represent the two sublattice spins, $n=6$ is the count of neighboring spins, $J$ and $D $ are the symmetric exchange and antisymmetric exchange constants, respectively, and $K_a$ and $K_c$ are the magnetic anisotropy components aligned with the $a$ and $c$ axes, respectively. $K_c$ is temperature-dependent to account for the spin reorientation transition. The magnon frequencies at the zone center are determined as:
\begin{align}
	  \hbar \Omega_{\sigma, qFM} &= 2S\left[ n(J+K_a)(K_a-K_c)\right ] ^{\frac{1}{2}}\\
	  \hbar \Omega_{\gamma, qAFM} &= S\left[ 4nJK_a+4K_a(K_a-K_c)+n^2D^2\right] ^{\frac{1}{2}}.
\end{align}
These expressions are used to fit the experimentally observed magnon frequencies by adopting reported values of $J$ and $D $ \cite{yamaguchi2013terahertz}. All parameter values are reported in Table S1 in the Supplementary Information. The fits along with the experimental magnon frequencies are plotted as a function of temperature in Fig. 2D. Although this model does not account for the redshift of the qAFM mode frequency toward the N\'{e}el transition temperature, it aligns well with the qFM mode frequencies across all temperatures. Notably, at about 200 K, the qAFM mode frequency matches twice the qFM mode frequency, i.e., $\Omega_{qAFM}=2\Omega_{qFM}$, serving as a benchmark for the parametric resonance condition. 
	
Having established the magnon responses in the linear response regime across a wide range of temperatures, we now examine the possibility for nonlinear parametric amplification of magnon coherences. We use a pair of nearly identical, linearly polarized, high-field THz pulses, each carrying a peak magnetic field strength of approximately 0.17 T, which interact sequentially with the sample. The magnetic field components of these THz fields are oriented 45$^{\circ}$ relative to the $a$ and $b$ crystallographic axes, ensuring the excitation of both magnon modes. The WGP is adjusted to exclusively select the THz magnetic field emission along the $a$ axis. As a result, only signals bearing a non-zero emission along the qFM mode axis are detectable, as depicted in Fig. 3A. By varying the inter-pulse delay time $\tau$ and recording the time-dependent signal fields induced by either or both THz pulses—leveraging a state-of-the-art single-shot detection technique\cite{teo2015invited,gao2022high}—we can isolate the nonlinear signals from coherent magnon emissions, i.e., $S(\tau,t)$. A subsequent 2D Fourier transformation of these nonlinear responses allows for the identification of signals originating from distinct excitation pathways\cite{junginger2012nonperturbative,woerner2013ultrafast,lu2016nonlinear,lu2017coherent,johnson2019distinguishing,houver20192d,mahmood2021observation,mashkovich2021terahertz,lin2022mapping,blank2023empowering,huang2024extreme}, which manifest as distinct peaks in the 2D frequency-frequency correlation maps, $S(\nu,f)$.

Fig. 3B showcases the 2D THz spectra obtained at a few selected temperatures, namely, 150 K, 200 K, and 300 K. These temperatures correspond to conditions below, at, and above the parametric resonance, with magnon frequencies: $\Omega_{qAFM}>2\Omega_{qFM}$, $\Omega_{qAFM}=2\Omega_{qFM}$, and $\Omega_{qAFM}<2\Omega_{qFM}$, respectively. Although these spectra feature multiple peaks, our analysis focuses primarily on peaks with the excitation frequency equal to the qAFM mode frequency, i.e., $\nu=\Omega_{qAFM}$. Divergence from the parametric resonance condition results in the appearance of two distinct peaks in the data. The detection frequencies of these peaks align with the qFM mode frequency, i.e., $f=\Omega_{qFM}$, and with the result of DFG from the two magnon modes, i.e., $f=\Omega_{qAFM}-\Omega_{qFM}$. The former peak is ascribed to a downconversion process wherein the first THz pulse excites the qAFM mode at frequency $\Omega_{qAFM}$, and this is succeeded by a secondary field interaction with frequency component at the difference frequency (i.e., $\Omega_{qAFM}-\Omega_{qFM}$), thus transferring the coherence from the qAFM mode to the qFM mode\cite{blank2023empowering}. The latter corresponds to the previously mentioned magnonic DFG resulting from magnon-magnon interaction after the first THz pulse excites the qAFM mode and the second excites the qFM mode\cite{zhang2024coupling}. Indeed, field-dependence measurements at 300 K show that the amplitudes of both peaks scale quadratically with the incident THz field strength (see Fig. 3C), and therefore both signals originate from second-order nonlinear processes. Thus, contributions from higher-order electric-field-driven processes are excluded. Furthermore, these signals vanish when the WGP is rotated by 90$^{\circ}$ to select the THz magnetic field emission along the $c$ axis (see Supplementary Fig. S2). This observation confirms that the DFG response shares the same symmetry as the qFM magnon emission and acts as a precursor to the degenerate parametric amplification. Indeed, at 200 K, the DFG frequency coincides with the qFM mode frequency, resulting in a single strong peak, indicative of degenerate parametric amplification.

For more quantitative analysis, we select a line cut across the excitation frequency of the qAFM mode, i.e., $\nu=\Omega_{qAFM}$, and represent the resulting spectra across a wide temperature range in Fig. 4A. Below 200 K, when $\Omega_{qAFM}>2\Omega_{qFM}$, the downconversion peak has a frequency lower than that of the DFG peak. Conversely, at temperatures exceeding 200 K, the frequency of the DFG peak is lower. Precisely at 200 K, which coincides with the parametric resonance condition, the two peaks converge to form a singular peak with substantially enhanced spectral amplitude. These features are confirmed in Fig. 4B, which depicts the temperature-dependent amplitude evolution of the downconversion peak, i.e., $[\Omega_{qFM},\Omega_{qAFM}]$. It is important to note that the downconversion signal is not only attributed to the THz-field-assisted process previously mentioned but also includes contributions from degenerate parametric amplification, occurring when the magnonic DFG resonates with the qFM mode. The data clearly show that the downconversion signal persists across all examined temperatures, peaking at 200 K, in accordance with the parametric resonance condition. Notably, this enhancement cannot be explained by the linear superposition of the downconversion and DFG signals, as the peak amplitude at 200 K far exceeds the sum of these signals at temperatures away from 200 K. This observation is corroborated by numerical solutions of the nonlinear equations of motion that describe the spin dynamics of the system, and the results are fitted with a model that integrates both the field-driven downconversion mechanism and the parametric amplification due to magnon-magnon interactions (see Supplementary Note 3). The simulation precisely reproduces the peak behavior as well as the general trend (See Fig. 4C). Notably, this pronounced peak indicates an enhanced magnon downconversion process, which can only be attributed to magnonic degenerate parametric amplification—a particular manifestation of magnon-magnon interactions—thus underscoring its significance to the observed phenomenon. Unlike the field-driven downconversion, degenerate parametric amplification continues even after the THz driving field ceases, as evidenced by the gradual increase in signal amplitude shown in Fig. S3D. Although these observations bear similarity to the previously reported magnon upconversion process, which also shows enhancement at the parametric resonance condition, they represent distinct nonlinear pathways: the upconversion process involves only driving the qFM mode to couple to the qAFM mode, while degenerate parametric amplification requires excitation of both the qAFM and qFM modes, analogous to its optical counterpart involving both pump and signal waves.

Our spectroscopic measurements unveil a crossover from the magnonic DFG to the degenerate parametric amplification of two distinct magnon modes in an antiferromagnet. The parametric amplification identified in this study is second-order relative to the THz fields and becomes significant when the frequency-matching condition is satisfied. This behavior contrasts with previous observations in ferromagnets\cite{bracher2017parallel,lisenkov2019magnetoelastic}, ferrimagnets\cite{van1967parametric}, and synthetic antiferromagnets\cite{kamimaki2020parametric} subjected to microwave and optical excitations, where the amplification occurs only after surpassing a specific instability threshold. As such, our method heralds a novel approach for the nonlinear coherent control of magnon modes in antiferromagnets, which are foundational to the design of magnonic oscillators and amplifiers. Analogous to the degenerate parametric amplification of light, the stimulated parametric down-conversion of antiferromagnetic magnons opens up new avenues for generating exotic magnonic states, such as magnon squeezing\cite{kamra2020magnon} and Bose-Einstein condensation of magnons\cite{demokritov2006bose}. In many antiferromagnets, the magnon frequencies are sensitive not only to temperature but also to other external parameters, including magnetic field, strain, and cavity engineering, making the prospects for exploitation of parametric amplification generally available, even at room temperature. More broadly, although parametric amplification of collective excitations has been proposed or observed in materials for phonons\cite{cartella2018parametric,juraschek2020parametric,haque2024terahertz}, Josephson plasmons\cite{von2022amplification}, and charge-density-wave amplitude modes\cite{michael2022optical}, our observation represents a rare example where parametric amplification appears at the zone center with two distinct modes involved. Therefore, the experimental methodology and the concept of stimulated parametric amplification are applicable to a diverse range of quantum materials, providing insight into nonlinear mode-mode interactions that are often overlooked in equilibrium states.

\newpage
\noindent\textbf{Acknowledgments} 
Z.Z., Z.-J.L., M.T.W. and K.A.N acknowledge support from the U.S. Department of Energy, Office of Basic Energy Sciences, under Award No. DE-SC0019126. Work at UT Austin was primarily supported by the Robert A. Welch Foundation (F-2092-20220331) (to F.Y.G. for setup building) and the United States Army Research Office (W911NF-23-1-0394) (to E.B. for data interpretation, manuscript writing and supervision). S.C. acknowledges the support from the National Natural Science Foundation of China (No. 12374116). Y.-C.C. acknowledges direct funding from the MIT UROP.

\noindent\textbf{Author contributions} Z.Z. conceived the project and performed experiments. Z.Z. analyzed the data with assistance from Y.C.-C. and M.T.W.. Z.Z. and Y.-C.C. performed the simulation and the theoretical analysis. Z.Z. and F.Y.G.  built the setup, assisted by Z.-J.L.. X. M. and S. C. provided the sample. Z.Z., E.B. and K.A.N. wrote the paper with input from all authors. K.A.N. and E.B. supervised the research.

\noindent\textbf{Competing interests} The authors declare no competing interests.

\noindent\textbf{Data availability}
Source data are provided with this paper. All other data that support the findings of this study are available from the corresponding authors on reasonable request.

\noindent\textbf{Code availability}
The codes used to perform the simulations and to analyse the data in this work are available from the corresponding authors upon request.

\section*{Methods}
\noindent \textit{Synthesis of ErFeO$_3$ single crystal} \\
A single crystal of ErFeO$_3$ (1.5 mm thick) grown by a floating zone melting technique was used in this work. The crystal was cut perpendicular to the $b$ axis. The crystallographic axes were determined by Laue diffraction measurements. The detailed synthesis procedure and characterization has been reported previously\cite{zhang2024terahertz}. 

\noindent \textit{Time-domain THz spectroscopy and two-dimensional THz spectroscopy} \\
The majority of the output of a 1 kHz Ti:Sapphire laser amplifier (Coherent Legend Elite Duo, 800 nm, 12 mJ, 35 fs) was split into two identical arms. These beams were then recombined with a controlled relative time delay to generate a pair of single-cycle THz pulses via tilted pulse front optical rectification in Mg:LiNbO$_3$\cite{yeh2007generation}. The THz beams were then directed onto a sample using two off-axis parabolic mirrors arranged in a $4f$ configuration. The transmitted THz light was then recollimated and refocused by another pair of $4f$ parabolic mirrors onto a 2-mm thick ZnTe detection crystal where it was overlapped with the electro-optic sampling probe beam derived from a small portion of the fundamental laser power. A single-shot detection method was employed to improve the signal-to-noise ratio by permitting the full time-dependence of the THz signal within a 20-ps window to be measured on a shot-to-shot basis.\cite{gao2022high,dastrup2024optical} In time-domain THz spectroscopy experiments, one of the optical pulses was blocked while the remaining THz pulse was attenuated by a pair of wire-grid polarizers to measure the primary linear response. For 2D THz measurements, differential chopping of the two pump beams ($A$ and $B$) was used to extract the nonlinear signal as the inter-pulse time $\tau$ was scanned across the desirable range\cite{gao2022high,lu2017coherent}. The detected THz signals were kept within the linear response regime of the detection crystal, as verified by attenuating the signals and seeing a linear dependence of the measured signal on the field strength incident on the EO crystal and no change in the 2D spectrum; thus, the nonlinear signal originated solely from the response of the sample. Similar verification was conducted for our earlier measurements\cite{zhang2024terahertz,zhang2024coupling} as well. The nonlinear signal $\mathbf{H}_{NL}$ is extracted as
\begin{equation}
	\mathbf{H}_{NL}(\tau,t) = \mathbf{H}_{AB}(\tau,t)  - \mathbf{H}_A(\tau,t)  - \mathbf{H}_B(t)  + \mathbf{H}_0(t),
\end{equation}
where $\mathbf{H}_{AB}$, $\mathbf{H}_A$, $\mathbf{H}_B$, and $\mathbf{H}_0$ are the THz signals recorded with pump $A$, pump $B$, both pumps and no pumps, respectively. A 2D Fourier transform of the resulting time-domain signal yields the 2D THz spectrum.

\newpage
\section*{References}

\footnotesize
\bibliographystyle{naturemag}
\bibliography{paper}

\begin{thebibliography}{10}
\expandafter\ifx\csname url\endcsname\relax
  \def\url#1{\texttt{#1}}\fi
\expandafter\ifx\csname urlprefix\endcsname\relax\def\urlprefix{URL }\fi
\providecommand{\bibinfo}[2]{#2}
\providecommand{\eprint}[2][]{\url{#2}}

\bibitem{boyd2008nonlinear}
\bibinfo{author}{Boyd, R.~W.}, \bibinfo{author}{Gaeta, A.~L.} \&
  \bibinfo{author}{Giese, E.}
\newblock \bibinfo{title}{Nonlinear optics}.
\newblock \bibinfo{pages}{1097--1110} (\bibinfo{publisher}{Springer},
  \bibinfo{year}{2008}).

\bibitem{louisell1961quantum}
\bibinfo{author}{Louisell, W.}, \bibinfo{author}{Yariv, A.} \&
  \bibinfo{author}{Siegman, A.}
\newblock \bibinfo{title}{Quantum fluctuations and noise in parametric
  processes. {I.}}
\newblock \emph{\bibinfo{journal}{Physical Review}}
  \textbf{\bibinfo{volume}{124}}, \bibinfo{pages}{1646} (\bibinfo{year}{1961}).

\bibitem{burnham1970observation}
\bibinfo{author}{Burnham, D.~C.} \& \bibinfo{author}{Weinberg, D.~L.}
\newblock \bibinfo{title}{Observation of simultaneity in parametric production
  of optical photon pairs}.
\newblock \emph{\bibinfo{journal}{Physical Review Letters}}
  \textbf{\bibinfo{volume}{25}}, \bibinfo{pages}{84} (\bibinfo{year}{1970}).

\bibitem{raiford1974degenerate}
\bibinfo{author}{Raiford, M.}
\newblock \bibinfo{title}{Degenerate parametric amplification with
  time-dependent pump amplitude and phase}.
\newblock \emph{\bibinfo{journal}{Physical Review A}}
  \textbf{\bibinfo{volume}{9}}, \bibinfo{pages}{2060} (\bibinfo{year}{1974}).

\bibitem{yuen1978optical}
\bibinfo{author}{Yuen, H.} \& \bibinfo{author}{Shapiro, J.}
\newblock \bibinfo{title}{Optical communication with two-photon coherent
  states--part i: Quantum-state propagation and quantum-noise}.
\newblock \emph{\bibinfo{journal}{IEEE Transactions on Information Theory}}
  \textbf{\bibinfo{volume}{24}}, \bibinfo{pages}{657--668}
  (\bibinfo{year}{1978}).

\bibitem{radic2008parametric}
\bibinfo{author}{Radic, S.}
\newblock \bibinfo{title}{Parametric amplification and processing in optical
  fibers}.
\newblock \emph{\bibinfo{journal}{Laser \& Photonics Reviews}}
  \textbf{\bibinfo{volume}{2}}, \bibinfo{pages}{498--513}
  (\bibinfo{year}{2008}).

\bibitem{marhic2015fiber}
\bibinfo{author}{Marhic, M.~E.} \emph{et~al.}
\newblock \bibinfo{title}{Fiber optical parametric amplifiers in optical
  communication systems}.
\newblock \emph{\bibinfo{journal}{Laser \& Photonics Reviews}}
  \textbf{\bibinfo{volume}{9}}, \bibinfo{pages}{50--74} (\bibinfo{year}{2015}).

\bibitem{milburn1981production}
\bibinfo{author}{Milburn, G.} \& \bibinfo{author}{Walls, D.}
\newblock \bibinfo{title}{Production of squeezed states in a degenerate
  parametric amplifier}.
\newblock \emph{\bibinfo{journal}{Optics Communications}}
  \textbf{\bibinfo{volume}{39}}, \bibinfo{pages}{401--404}
  (\bibinfo{year}{1981}).

\bibitem{yamamoto1986preparation}
\bibinfo{author}{Yamamoto, Y.} \& \bibinfo{author}{Haus, H.}
\newblock \bibinfo{title}{Preparation, measurement and information capacity of
  optical quantum states}.
\newblock \emph{\bibinfo{journal}{Reviews of Modern Physics}}
  \textbf{\bibinfo{volume}{58}}, \bibinfo{pages}{1001} (\bibinfo{year}{1986}).

\bibitem{eichler2014quantum}
\bibinfo{author}{Eichler, C.}, \bibinfo{author}{Salathe, Y.},
  \bibinfo{author}{Mlynek, J.}, \bibinfo{author}{Schmidt, S.} \&
  \bibinfo{author}{Wallraff, A.}
\newblock \bibinfo{title}{Quantum-limited amplification and entanglement in
  coupled nonlinear resonators}.
\newblock \emph{\bibinfo{journal}{Physical Review Letters}}
  \textbf{\bibinfo{volume}{113}}, \bibinfo{pages}{110502}
  (\bibinfo{year}{2014}).

\bibitem{ourjoumtsev2006generating}
\bibinfo{author}{Ourjoumtsev, A.}, \bibinfo{author}{Tualle-Brouri, R.},
  \bibinfo{author}{Laurat, J.} \& \bibinfo{author}{Grangier, P.}
\newblock \bibinfo{title}{Generating optical schrodinger kittens for quantum
  information processing}.
\newblock \emph{\bibinfo{journal}{Science}} \textbf{\bibinfo{volume}{312}},
  \bibinfo{pages}{83--86} (\bibinfo{year}{2006}).

\bibitem{bergeal2010phase}
\bibinfo{author}{Bergeal, N.} \emph{et~al.}
\newblock \bibinfo{title}{Phase-preserving amplification near the quantum limit
  with a josephson ring modulator}.
\newblock \emph{\bibinfo{journal}{Nature}} \textbf{\bibinfo{volume}{465}},
  \bibinfo{pages}{64--68} (\bibinfo{year}{2010}).

\bibitem{pirro2021advances}
\bibinfo{author}{Pirro, P.}, \bibinfo{author}{Vasyuchka, V.~I.},
  \bibinfo{author}{Serga, A.~A.} \& \bibinfo{author}{Hillebrands, B.}
\newblock \bibinfo{title}{Advances in coherent magnonics}.
\newblock \emph{\bibinfo{journal}{Nature Reviews Materials}}
  \textbf{\bibinfo{volume}{6}}, \bibinfo{pages}{1114--1135}
  (\bibinfo{year}{2021}).

\bibitem{zheng2023tutorial}
\bibinfo{author}{Zheng, S.} \emph{et~al.}
\newblock \bibinfo{title}{Tutorial: Nonlinear magnonics}.
\newblock \emph{\bibinfo{journal}{Journal of Applied Physics}}
  \textbf{\bibinfo{volume}{134}}, \bibinfo{pages}{151101}
  (\bibinfo{year}{2023}).

\bibitem{zhang2024terahertz}
\bibinfo{author}{Zhang, Z.} \emph{et~al.}
\newblock \bibinfo{title}{Terahertz-field-driven magnon upconversion in an
  antiferromagnet}.
\newblock \emph{\bibinfo{journal}{Nature Physics}} \bibinfo{pages}{1--6}
  (\bibinfo{year}{2024}).

\bibitem{zhang2024coupling}
\bibinfo{author}{Zhang, Z.} \emph{et~al.}
\newblock \bibinfo{title}{Terahertz field-induced nonlinear coupling of two
  magnon modes in an antiferromagnet}.
\newblock \emph{\bibinfo{journal}{Nature Physics}} \bibinfo{pages}{1--6}
  (\bibinfo{year}{2024}).

\bibitem{leenders2024canted}
\bibinfo{author}{Leenders, R.}, \bibinfo{author}{Afanasiev, D.},
  \bibinfo{author}{Kimel, A.} \& \bibinfo{author}{Mikhaylovskiy, R.}
\newblock \bibinfo{title}{Canted spin order as a platform for ultrafast
  conversion of magnons}.
\newblock \emph{\bibinfo{journal}{Nature}} \bibinfo{pages}{1--5}
  (\bibinfo{year}{2024}).

\bibitem{bracher2017parallel}
\bibinfo{author}{Br{\"a}cher, T.}, \bibinfo{author}{Pirro, P.} \&
  \bibinfo{author}{Hillebrands, B.}
\newblock \bibinfo{title}{Parallel pumping for magnon spintronics:
  Amplification and manipulation of magnon spin currents on the micron-scale}.
\newblock \emph{\bibinfo{journal}{Physics Reports}}
  \textbf{\bibinfo{volume}{699}}, \bibinfo{pages}{1--34}
  (\bibinfo{year}{2017}).

\bibitem{lisenkov2019magnetoelastic}
\bibinfo{author}{Lisenkov, I.}, \bibinfo{author}{Jander, A.} \&
  \bibinfo{author}{Dhagat, P.}
\newblock \bibinfo{title}{Magnetoelastic parametric instabilities of localized
  spin waves induced by traveling elastic waves}.
\newblock \emph{\bibinfo{journal}{Physical Review B}}
  \textbf{\bibinfo{volume}{99}}, \bibinfo{pages}{184433}
  (\bibinfo{year}{2019}).

\bibitem{nvemec2018antiferromagnetic}
\bibinfo{author}{N{\v{e}}mec, P.}, \bibinfo{author}{Fiebig, M.},
  \bibinfo{author}{Kampfrath, T.} \& \bibinfo{author}{Kimel, A.~V.}
\newblock \bibinfo{title}{Antiferromagnetic opto-spintronics}.
\newblock \emph{\bibinfo{journal}{Nature Physics}}
  \textbf{\bibinfo{volume}{14}}, \bibinfo{pages}{229--241}
  (\bibinfo{year}{2018}).

\bibitem{rezende2019introduction}
\bibinfo{author}{Rezende, S.~M.}, \bibinfo{author}{Azevedo, A.} \&
  \bibinfo{author}{Rodr{\'\i}guez-Su{\'a}rez, R.~L.}
\newblock \bibinfo{title}{Introduction to antiferromagnetic magnons}.
\newblock \emph{\bibinfo{journal}{Journal of Applied Physics}}
  \textbf{\bibinfo{volume}{126}} (\bibinfo{year}{2019}).

\bibitem{han2023coherent}
\bibinfo{author}{Han, J.}, \bibinfo{author}{Cheng, R.}, \bibinfo{author}{Liu,
  L.}, \bibinfo{author}{Ohno, H.} \& \bibinfo{author}{Fukami, S.}
\newblock \bibinfo{title}{Coherent antiferromagnetic spintronics}.
\newblock \emph{\bibinfo{journal}{Nature Materials}}
  \textbf{\bibinfo{volume}{22}}, \bibinfo{pages}{684--695}
  (\bibinfo{year}{2023}).

\bibitem{huang2024extreme}
\bibinfo{author}{Huang, C.} \emph{et~al.}
\newblock \bibinfo{title}{Extreme terahertz magnon multiplication induced by
  resonant magnetic pulse pairs}.
\newblock \emph{\bibinfo{journal}{Nature Communications}}
  \textbf{\bibinfo{volume}{15}}, \bibinfo{pages}{3214} (\bibinfo{year}{2024}).

\bibitem{li2023terahertz}
\bibinfo{author}{Li, X.}, \bibinfo{author}{Kim, D.}, \bibinfo{author}{Liu, Y.}
  \& \bibinfo{author}{Kono, J.}
\newblock \bibinfo{title}{Terahertz spin dynamics in rare-earth orthoferrites}.
\newblock \emph{\bibinfo{journal}{Photonics Insights}}
  \textbf{\bibinfo{volume}{1}}, \bibinfo{pages}{R05--R05}
  (\bibinfo{year}{2023}).

\bibitem{li2022perspective}
\bibinfo{author}{Li, J.}, \bibinfo{author}{Yang, C.-J.},
  \bibinfo{author}{Mondal, R.}, \bibinfo{author}{Tzschaschel, C.} \&
  \bibinfo{author}{Pal, S.}
\newblock \bibinfo{title}{A perspective on nonlinearities in coherent
  magnetization dynamics}.
\newblock \emph{\bibinfo{journal}{Applied Physics Letters}}
  \textbf{\bibinfo{volume}{120}} (\bibinfo{year}{2022}).

\bibitem{lu2017coherent}
\bibinfo{author}{Lu, J.} \emph{et~al.}
\newblock \bibinfo{title}{Coherent two-dimensional terahertz magnetic resonance
  spectroscopy of collective spin waves}.
\newblock \emph{\bibinfo{journal}{Physical Review Letters}}
  \textbf{\bibinfo{volume}{118}}, \bibinfo{pages}{207204}
  (\bibinfo{year}{2017}).

\bibitem{metzger2024magnon}
\bibinfo{author}{Metzger, T.~W.} \emph{et~al.}
\newblock \bibinfo{title}{Magnon-phonon fermi resonance in antiferromagnetic
  {CoF$_2$}}.
\newblock \emph{\bibinfo{journal}{Nature Communications}}
  \textbf{\bibinfo{volume}{15}}, \bibinfo{pages}{5472} (\bibinfo{year}{2024}).

\bibitem{yamaguchi2013terahertz}
\bibinfo{author}{Yamaguchi, K.}, \bibinfo{author}{Kurihara, T.},
  \bibinfo{author}{Minami, Y.}, \bibinfo{author}{Nakajima, M.} \&
  \bibinfo{author}{Suemoto, T.}
\newblock \bibinfo{title}{Terahertz time-domain observation of spin
  reorientation in orthoferrite {ErFeO$_3$} through magnetic free induction
  decay}.
\newblock \emph{\bibinfo{journal}{Physical Review Letters}}
  \textbf{\bibinfo{volume}{110}}, \bibinfo{pages}{137204}
  (\bibinfo{year}{2013}).

\bibitem{li2018observation}
\bibinfo{author}{Li, X.} \emph{et~al.}
\newblock \bibinfo{title}{Observation of {Dicke} cooperativity in magnetic
  interactions}.
\newblock \emph{\bibinfo{journal}{Science}} \textbf{\bibinfo{volume}{361}},
  \bibinfo{pages}{794--797} (\bibinfo{year}{2018}).

\bibitem{teo2015invited}
\bibinfo{author}{Teo, S.~M.}, \bibinfo{author}{Ofori-Okai, B.~K.},
  \bibinfo{author}{Werley, C.~A.} \& \bibinfo{author}{Nelson, K.~A.}
\newblock \bibinfo{title}{Invited article: Single-shot {THz} detection
  techniques optimized for multidimensional {THz} spectroscopy}.
\newblock \emph{\bibinfo{journal}{Review of Scientific Instruments}}
  \textbf{\bibinfo{volume}{86}} (\bibinfo{year}{2015}).

\bibitem{gao2022high}
\bibinfo{author}{Gao, F.~Y.}, \bibinfo{author}{Zhang, Z.},
  \bibinfo{author}{Liu, Z.-J.} \& \bibinfo{author}{Nelson, K.~A.}
\newblock \bibinfo{title}{High-speed two-dimensional terahertz spectroscopy
  with echelon-based shot-to-shot balanced detection}.
\newblock \emph{\bibinfo{journal}{Optics Letters}}
  \textbf{\bibinfo{volume}{47}}, \bibinfo{pages}{3479--3482}
  (\bibinfo{year}{2022}).

\bibitem{junginger2012nonperturbative}
\bibinfo{author}{Junginger, F.} \emph{et~al.}
\newblock \bibinfo{title}{Nonperturbative interband response of a bulk {InSb}
  semiconductor driven off resonantly by terahertz electromagnetic few-cycle
  pulses}.
\newblock \emph{\bibinfo{journal}{Physical Review Letters}}
  \textbf{\bibinfo{volume}{109}}, \bibinfo{pages}{147403}
  (\bibinfo{year}{2012}).

\bibitem{woerner2013ultrafast}
\bibinfo{author}{Woerner, M.}, \bibinfo{author}{Kuehn, W.},
  \bibinfo{author}{Bowlan, P.}, \bibinfo{author}{Reimann, K.} \&
  \bibinfo{author}{Elsaesser, T.}
\newblock \bibinfo{title}{Ultrafast two-dimensional terahertz spectroscopy of
  elementary excitations in solids}.
\newblock \emph{\bibinfo{journal}{New Journal of Physics}}
  \textbf{\bibinfo{volume}{15}}, \bibinfo{pages}{025039}
  (\bibinfo{year}{2013}).

\bibitem{lu2016nonlinear}
\bibinfo{author}{Lu, J.} \emph{et~al.}
\newblock \bibinfo{title}{Nonlinear two-dimensional terahertz photon echo and
  rotational spectroscopy in the gas phase}.
\newblock \emph{\bibinfo{journal}{Proceedings of the National Academy of
  Sciences}} \textbf{\bibinfo{volume}{113}}, \bibinfo{pages}{11800--11805}
  (\bibinfo{year}{2016}).

\bibitem{johnson2019distinguishing}
\bibinfo{author}{Johnson, C.~L.}, \bibinfo{author}{Knighton, B.~E.} \&
  \bibinfo{author}{Johnson, J.~A.}
\newblock \bibinfo{title}{Distinguishing nonlinear terahertz excitation
  pathways with two-dimensional spectroscopy}.
\newblock \emph{\bibinfo{journal}{Physical Review Letters}}
  \textbf{\bibinfo{volume}{122}}, \bibinfo{pages}{073901}
  (\bibinfo{year}{2019}).

\bibitem{houver20192d}
\bibinfo{author}{Houver, S.}, \bibinfo{author}{Huber, L.},
  \bibinfo{author}{Savoini, M.}, \bibinfo{author}{Abreu, E.} \&
  \bibinfo{author}{Johnson, S.~L.}
\newblock \bibinfo{title}{{2D THz} spectroscopic investigation of ballistic
  conduction-band electron dynamics in insb}.
\newblock \emph{\bibinfo{journal}{Optics Express}}
  \textbf{\bibinfo{volume}{27}}, \bibinfo{pages}{10854--10865}
  (\bibinfo{year}{2019}).

\bibitem{mahmood2021observation}
\bibinfo{author}{Mahmood, F.}, \bibinfo{author}{Chaudhuri, D.},
  \bibinfo{author}{Gopalakrishnan, S.}, \bibinfo{author}{Nandkishore, R.} \&
  \bibinfo{author}{Armitage, N.}
\newblock \bibinfo{title}{Observation of a marginal {Fermi} glass}.
\newblock \emph{\bibinfo{journal}{Nature Physics}}
  \textbf{\bibinfo{volume}{17}}, \bibinfo{pages}{627--631}
  (\bibinfo{year}{2021}).

\bibitem{mashkovich2021terahertz}
\bibinfo{author}{Mashkovich, E.~A.} \emph{et~al.}
\newblock \bibinfo{title}{Terahertz light--driven coupling of antiferromagnetic
  spins to lattice}.
\newblock \emph{\bibinfo{journal}{Science}} \textbf{\bibinfo{volume}{374}},
  \bibinfo{pages}{1608--1611} (\bibinfo{year}{2021}).

\bibitem{lin2022mapping}
\bibinfo{author}{Lin, H.-W.}, \bibinfo{author}{Mead, G.} \&
  \bibinfo{author}{Blake, G.~A.}
\newblock \bibinfo{title}{Mapping {LiNbO$_3$} phonon-polariton nonlinearities
  with 2d {THz-THz-Raman} spectroscopy}.
\newblock \emph{\bibinfo{journal}{Physical Review Letters}}
  \textbf{\bibinfo{volume}{129}}, \bibinfo{pages}{207401}
  (\bibinfo{year}{2022}).

\bibitem{blank2023empowering}
\bibinfo{author}{Blank, T.~G.} \emph{et~al.}
\newblock \bibinfo{title}{Empowering control of antiferromagnets by
  {THz}-induced spin coherence}.
\newblock \emph{\bibinfo{journal}{Physical Review Letters}}
  \textbf{\bibinfo{volume}{131}}, \bibinfo{pages}{096701}
  (\bibinfo{year}{2023}).

\bibitem{van1967parametric}
\bibinfo{author}{Van~de Vaart, H.}, \bibinfo{author}{Lyons, D.} \&
  \bibinfo{author}{Damon, R.}
\newblock \bibinfo{title}{Parametric excitation and amplification of
  magnetoelastic waves}.
\newblock \emph{\bibinfo{journal}{Journal of Applied Physics}}
  \textbf{\bibinfo{volume}{38}}, \bibinfo{pages}{360--374}
  (\bibinfo{year}{1967}).

\bibitem{kamimaki2020parametric}
\bibinfo{author}{Kamimaki, A.}, \bibinfo{author}{Iihama, S.},
  \bibinfo{author}{Suzuki, K.}, \bibinfo{author}{Yoshinaga, N.} \&
  \bibinfo{author}{Mizukami, S.}
\newblock \bibinfo{title}{Parametric amplification of magnons in synthetic
  antiferromagnets}.
\newblock \emph{\bibinfo{journal}{Physical Review Applied}}
  \textbf{\bibinfo{volume}{13}}, \bibinfo{pages}{044036}
  (\bibinfo{year}{2020}).

\bibitem{kamra2020magnon}
\bibinfo{author}{Kamra, A.}, \bibinfo{author}{Belzig, W.} \&
  \bibinfo{author}{Brataas, A.}
\newblock \bibinfo{title}{Magnon-squeezing as a niche of quantum magnonics}.
\newblock \emph{\bibinfo{journal}{Applied Physics Letters}}
  \textbf{\bibinfo{volume}{117}} (\bibinfo{year}{2020}).

\bibitem{demokritov2006bose}
\bibinfo{author}{Demokritov, S.~O.} \emph{et~al.}
\newblock \bibinfo{title}{Bose--einstein condensation of quasi-equilibrium
  magnons at room temperature under pumping}.
\newblock \emph{\bibinfo{journal}{Nature}} \textbf{\bibinfo{volume}{443}},
  \bibinfo{pages}{430--433} (\bibinfo{year}{2006}).

\bibitem{cartella2018parametric}
\bibinfo{author}{Cartella, A.}, \bibinfo{author}{Nova, T.~F.},
  \bibinfo{author}{Fechner, M.}, \bibinfo{author}{Merlin, R.} \&
  \bibinfo{author}{Cavalleri, A.}
\newblock \bibinfo{title}{Parametric amplification of optical phonons}.
\newblock \emph{\bibinfo{journal}{Proceedings of the National Academy of
  Sciences}} \textbf{\bibinfo{volume}{115}}, \bibinfo{pages}{12148--12151}
  (\bibinfo{year}{2018}).

\bibitem{juraschek2020parametric}
\bibinfo{author}{Juraschek, D.~M.}, \bibinfo{author}{Meier, Q.~N.} \&
  \bibinfo{author}{Narang, P.}
\newblock \bibinfo{title}{Parametric excitation of an optically silent
  {Goldstone}-like phonon mode}.
\newblock \emph{\bibinfo{journal}{Physical Review Letters}}
  \textbf{\bibinfo{volume}{124}}, \bibinfo{pages}{117401}
  (\bibinfo{year}{2020}).

\bibitem{haque2024terahertz}
\bibinfo{author}{Haque, S. R.~U.} \emph{et~al.}
\newblock \bibinfo{title}{Terahertz parametric amplification as a reporter of
  exciton condensate dynamics}.
\newblock \emph{\bibinfo{journal}{Nature Materials}} \bibinfo{pages}{1--7}
  (\bibinfo{year}{2024}).

\bibitem{von2022amplification}
\bibinfo{author}{von Hoegen, A.} \emph{et~al.}
\newblock \bibinfo{title}{Amplification of superconducting fluctuations in
  driven {YBa}$_2${Cu}$_3${O}$_{6+x}$}.
\newblock \emph{\bibinfo{journal}{Physical Review X}}
  \textbf{\bibinfo{volume}{12}}, \bibinfo{pages}{031008}
  (\bibinfo{year}{2022}).

\bibitem{michael2022optical}
\bibinfo{author}{Michael, S.} \& \bibinfo{author}{Schneider, H.~C.}
\newblock \bibinfo{title}{Optical amplification in a charge density wave phase
  of a quasi-two-dimensional material}.
\newblock \emph{\bibinfo{journal}{Physical Review B}}
  \textbf{\bibinfo{volume}{105}}, \bibinfo{pages}{235108}
  (\bibinfo{year}{2022}).

\bibitem{yeh2007generation}
\bibinfo{author}{Yeh, K.-L.}, \bibinfo{author}{Hoffmann, M.},
  \bibinfo{author}{Hebling, J.} \& \bibinfo{author}{Nelson, K.~A.}
\newblock \bibinfo{title}{Generation of 10 {$\mu$J} ultrashort terahertz pulses
  by optical rectification}.
\newblock \emph{\bibinfo{journal}{Applied Physics Letters}}
  \textbf{\bibinfo{volume}{90}} (\bibinfo{year}{2007}).

\bibitem{dastrup2024optical}
\bibinfo{author}{Dastrup, B.~S.}, \bibinfo{author}{Miedaner, P.~R.},
  \bibinfo{author}{Zhang, Z.} \& \bibinfo{author}{Nelson, K.~A.}
\newblock \bibinfo{title}{Optical-pump--terahertz-probe spectroscopy in high
  magnetic fields with {kHz} single-shot detection}.
\newblock \emph{\bibinfo{journal}{Review of Scientific Instruments}}
  \textbf{\bibinfo{volume}{95}} (\bibinfo{year}{2024}).

\end{thebibliography}

\newpage
\FloatBarrier
\begin{figure}
	\centering
	\includegraphics[width=0.95\linewidth]{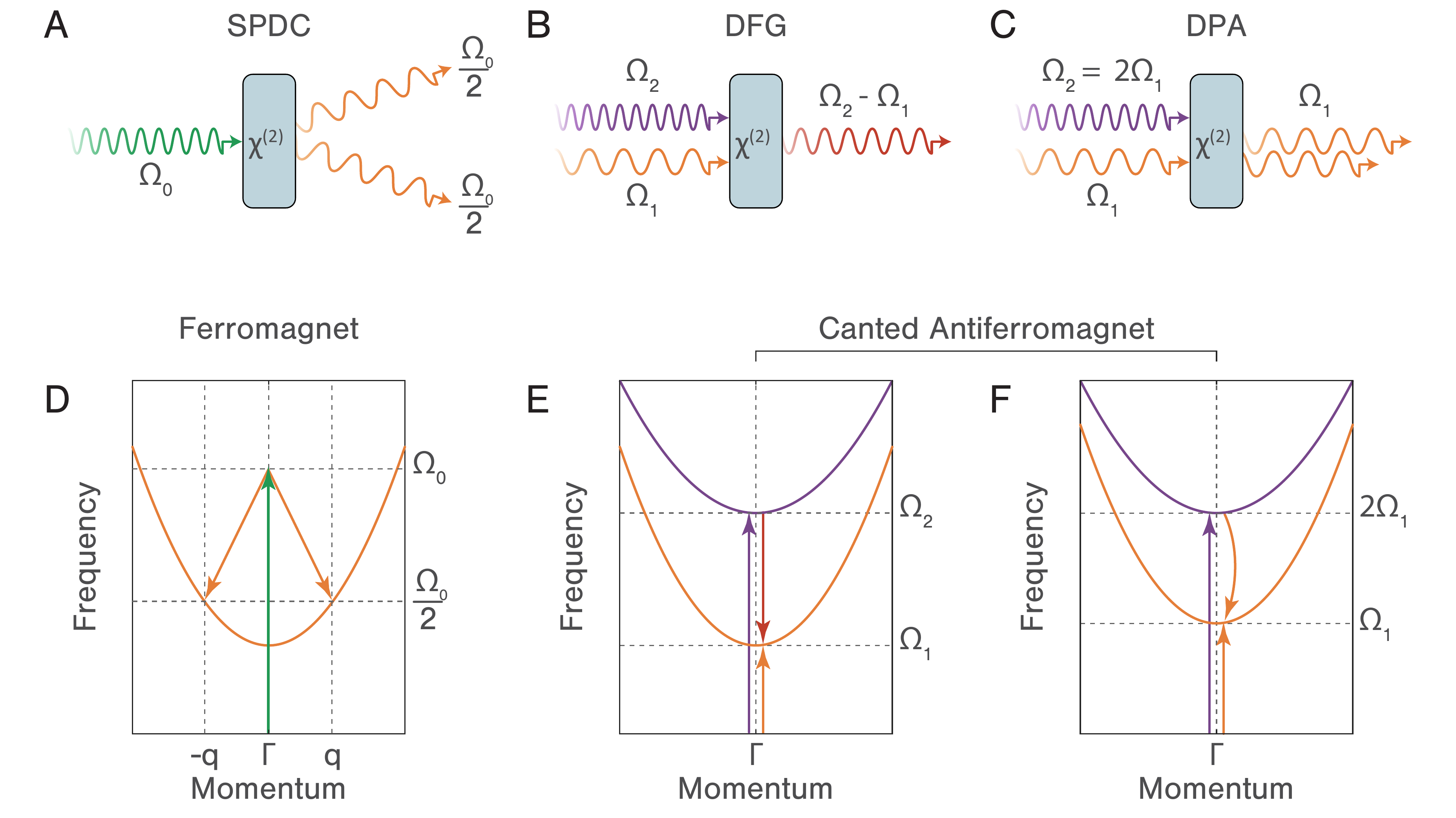}
	\caption{\label{fig:Fig1}
	Nonlinear optics and nonlinear magnonics. (A) Spontaneous parametric down-conversion (SPDC). A pump wave with frequency $\Omega_0$ generates a pair of optical photons at half the pump frequency (i.e., $\frac{\Omega_0}{2}$), conserving both momentum and energy.  (B) Difference frequency generation (DFG). A signal wave with frequency $\Omega_1$ and a pump wave with frequency $\Omega_2$ generate an idler wave at frequency $\Omega_2-\Omega_1$. (C) Degenerate parametric amplification (DPA). A signal wave with frequency $\Omega_1$ and a pump wave with frequency $\Omega_2=2\Omega_1$ result in the amplification of the signal wave. (D) Parametric excitation of magnon modes in a ferromagnet. A strong microwave field with frequency $\Omega_0$ directly couples to two counter-propagating spin waves at wavevectors $\pm q$ with half of the pump frequency $\frac{\Omega_0}{2}$. (E) DFG of two distinct magnon modes in a canted antiferromagnet. Coherent excitation of both magnon modes leads to coherent photon emission at frequency $\Omega_2-\Omega_1$. (F) Degenerate parametric amplification of a coherent magnon mode in a canted antiferromagnet. When $\Omega_2=2\Omega_1$, driving both magnon modes leads to the amplification of the lower frequency mode.
}
\end{figure}

\begin{figure}
	\centering
	\includegraphics[width=0.95\linewidth]{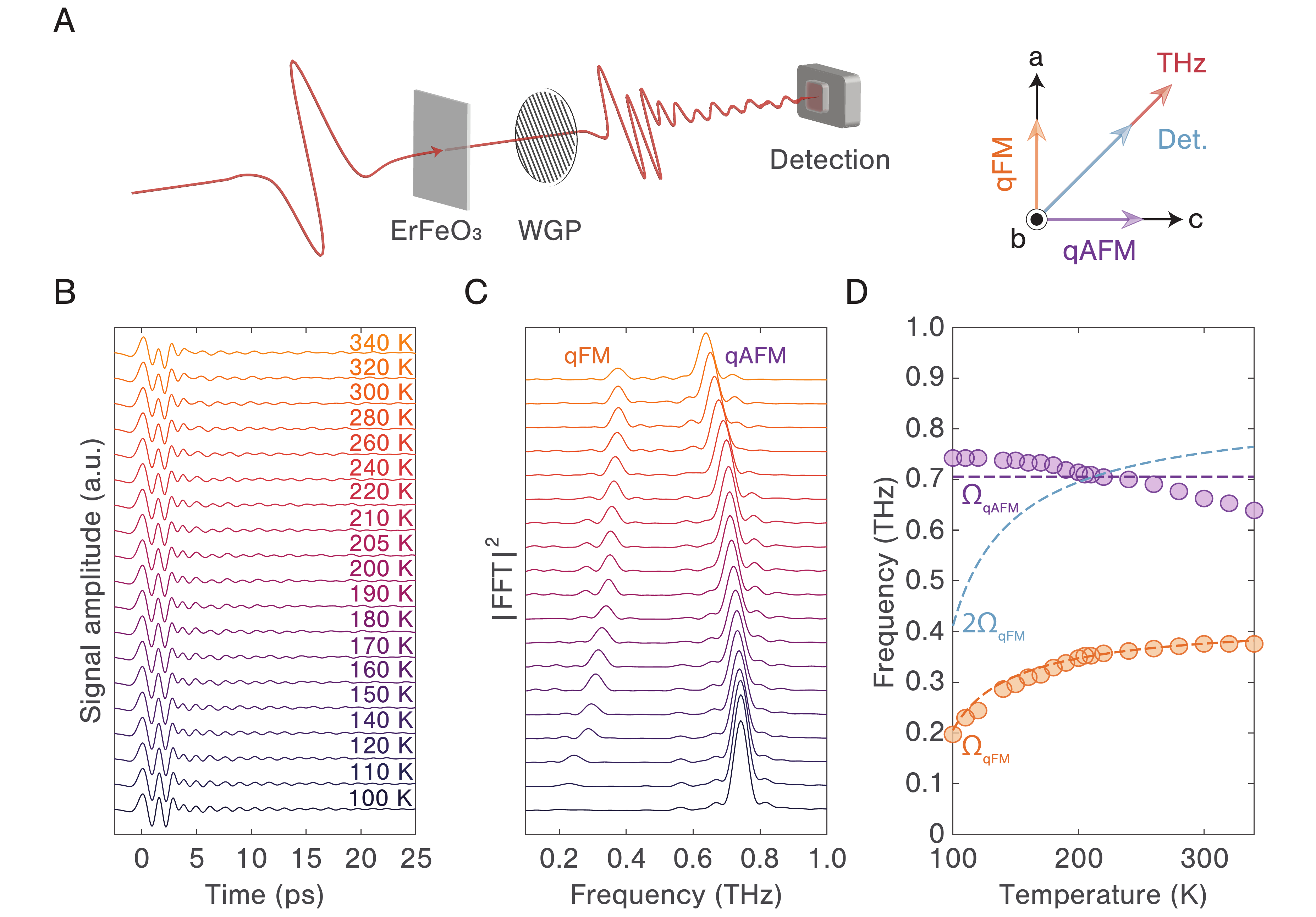}
	\caption{\label{fig:Fig2}  
	Temperature-dependent linear magnon response. (A) Sketch of the experimental setup for time-domain THz spectroscopy measurements. Both the polarization of the THz pulse and the WGP are set at a 45$^{\circ}$ angle relative to both $a$ and $c$ crystallographic axes. (B) Time-domain THz signals showing the excitation of both magnon modes across a range of temperatures. The initial double-peak profile in the primary THz signals arises from a temporal walk-off of the THz field components along different crystallographic axes. (C) Fourier transforms of the oscillatory responses following the primary THz peaks presented in (B). (D) Experimentally derived frequencies for both the qFM (orange dots) and qAFM (purple dots) magnon modes plotted against temperature. Fits are based on the uniform two-spin model, represented by dashed lines in corresponding colors. The dashed blue line signifies double the frequency of the qFM mode, which intersects with the qAFM mode frequency at approximately 200 K.}
\end{figure}

\begin{figure}
	\centering
	\includegraphics[width=0.95\linewidth]{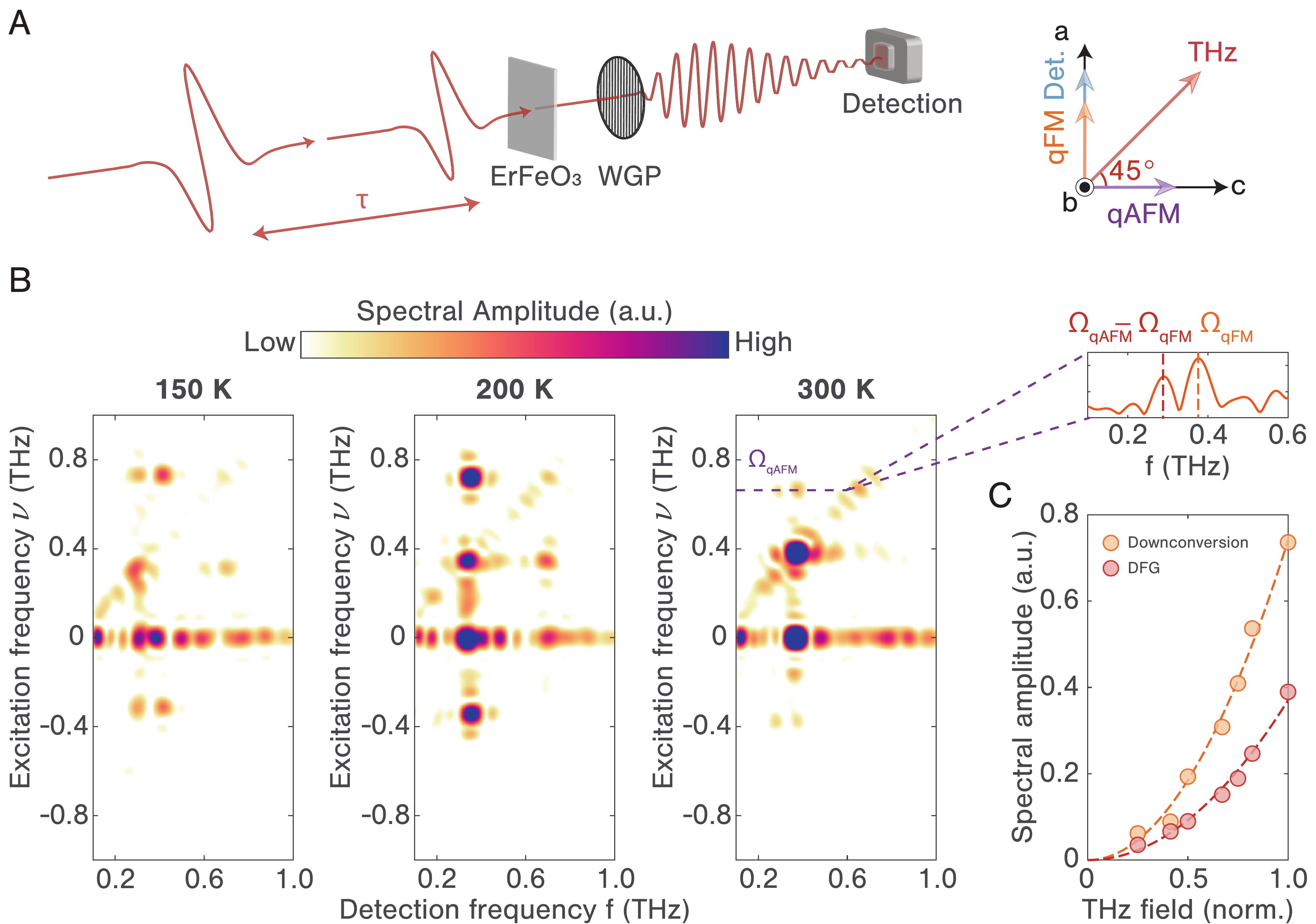}
	\caption{\label{fig:Fig3}  
	Polarization-selective 2D THz spectroscopy measurements. (A) Sketch of the experimental setup for the polarization-selective 2D THz spectroscopy measurements. The polarization of both THz pulses is set at a 45$^{\circ}$ angle relative to both $a$ and $c$ axes. The WGP is adjusted to allow for the detection of magnetization emissions along the $a$ axis, while rejecting those along the $c$ axis. (B) Representative 2D THz spectra at 150 K (left), 200 K (middle), and 300 K (right). For the 300 K data, a spectral line cut along the excitation frequency ${\nu=\Omega}_{qAFM}$, is shown. Detection frequencies are indicated by the dashed red line for $f=\Omega_{qAFM}-\Omega_{qFM}$ and the dashed orange line for $f=\Omega_{qFM}$. (C) THz magnetic field dependencies of the spectral amplitudes for both DFG (red dots) and downconversion (orange dots) signals at 300 K, accompanied by their respective quadratic fits (dashed lines).}
\end{figure}

\begin{figure}
	\centering
	\includegraphics[width=0.95\linewidth]{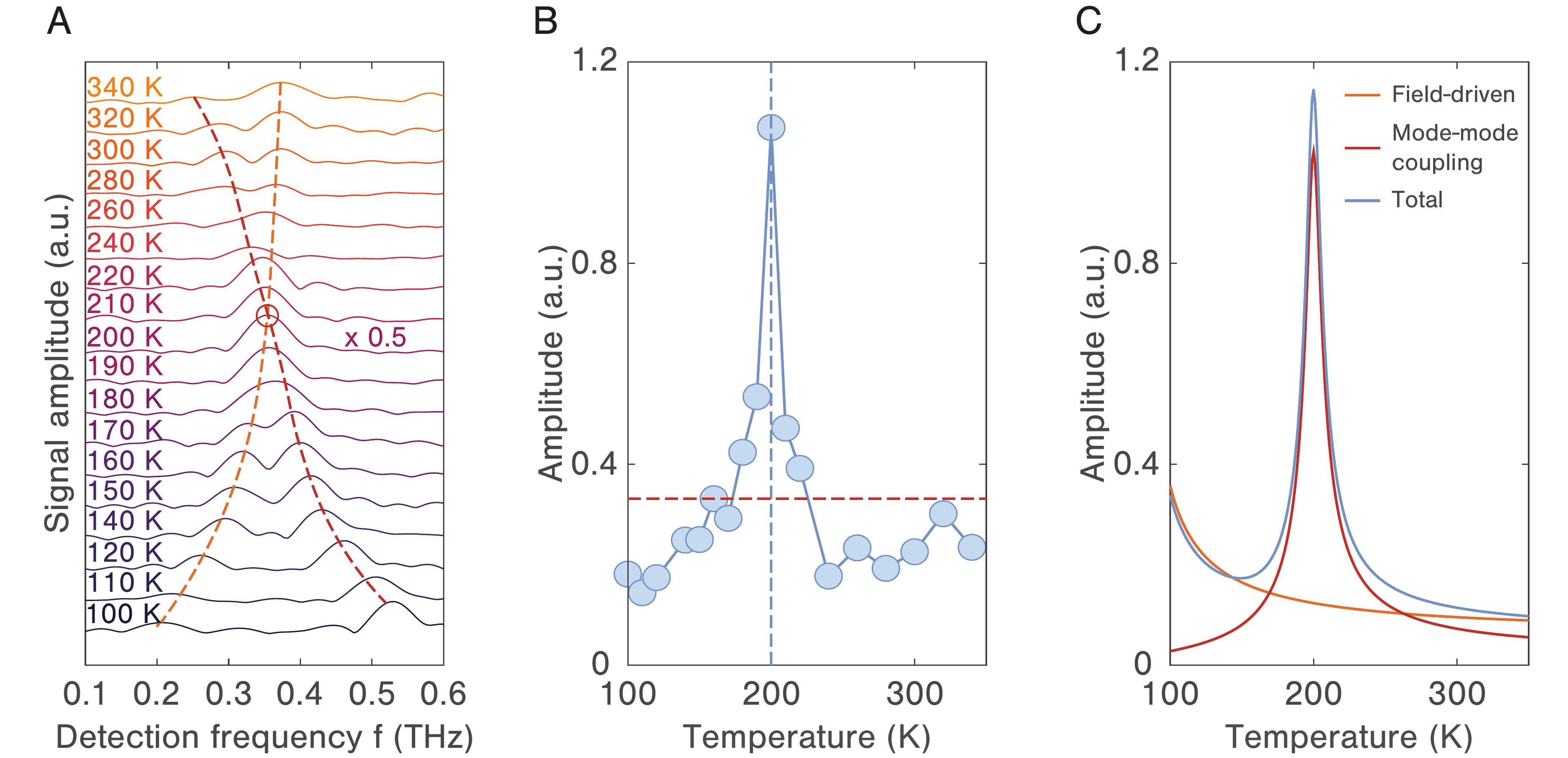}
	\caption{\label{fig:Fig4}  
	Stimulated parametric downconversion to the qFM mode. (A) Spectral line cuts along the excitation frequency ${\nu=\Omega}_{qAFM}$ across a broad range of temperatures. The spectral amplitude at 200 K is multiplied by 0.5 for better visualization. The dashed curves indicate the down-conversion frequencies (i.e., $\Omega_{qFM}$) increasing with temperature) and the DFG frequencies (decreasing with temperature). (B) Temperature dependence of the amplitude of the downconversion signal. The amplitude peaks at 200 K (blue dashed line), where the qAFM mode frequency matches twice the qFM mode frequency ($\Omega_{qAFM}={2\Omega}_{qFM}$). The red dashed line shows the averaged amplitude of the DFG signals at temperatures away from 200 K, which is smaller than half the peak amplitude at 200 K. (C) Simulated downconversion amplitude contributed by the field-driven mechanism (orange) and the parametric amplification mechanism due to magnon-magnon interactions (red line). The blue line indicates the sum of the two contributions.}
\end{figure}
\end{document}